\documentclass[aps,prb,twocolumn,superscriptaddress,showpacs,preprintnumbers]{revtex4-1}
\usepackage{graphicx }
\usepackage{dcolumn}
\usepackage{bm}
\usepackage{braket}
\usepackage{amsmath}
\usepackage[caption=false]{subfig} 
\usepackage{setspace} 
\usepackage{tabularx}
\usepackage{color}
\usepackage{epstopdf}
\newcolumntype{Y}{>{\centering\arraybackslash}X}

\newcommand{\tn}{\textit{T}$_{\mathrm{N}}$}

\newcommand{\ton}{\textit{T}$_{\mathrm{N}}^{\mathrm{onset}}$}
\newcommand{\tfull}{\textit{T}$_{\mathrm{N}}^{\mathrm{100 \%}}$}

\begin{document}

\preprint{insert preprint}

\title{Microscopic phase diagram of LaFeAsO single crystals under pressure}
\author{Philipp Materne}
\altaffiliation[]{pmaterne@anl.gov}
\affiliation{Argonne National Laboratory, Lemont, IL 60439, USA}
\author{Wenli Bi}
\affiliation{Department of Geology, University of Illinois at Urbana-Champaign, Urbana, Illinois 61801, USA}
\affiliation{Argonne National Laboratory, Lemont, IL 60439, USA}
\author{Jiyong Zhao}
\author{Michael Y. Hu}
\affiliation{Argonne National Laboratory, Lemont, IL 60439, USA}
\author{Rhea Kappenberger}
\affiliation{Leibniz Institute for Solid State and Materials Research (IFW) Dresden, D-01069, Germany}
\affiliation{Institute of Solid State and Materials Physics, TU Dresden, D-01069 Dresden, Germany}
\author{Sabine Wurmehl}
\affiliation{Leibniz Institute for Solid State and Materials Research (IFW) Dresden, D-01069, Germany}
\affiliation{Institute of Solid State and Materials Physics, TU Dresden, D-01069 Dresden, Germany}
\author{Saicharan Aswartham}
\affiliation{Leibniz Institute for Solid State and Materials Research (IFW) Dresden, D-01069, Germany}
\author{Bernd B\"uchner}
\affiliation{Leibniz Institute for Solid State and Materials Research (IFW) Dresden, D-01069, Germany}
\affiliation{Institute of Solid State and Materials Physics, TU Dresden, D-01069 Dresden, Germany}
\author{E. Ercan Alp}
\affiliation{Argonne National Laboratory, Lemont, IL 60439, USA}


\date{\today}

\begin{abstract}
We investigated LaFeAsO single crystals by means of synchrotron M\"ossbauer spectroscopy under pressure up to 7.5 GPa and down to 13 K and provide a microscopic phase diagram.
We found a continuous suppression of the magnetic hyperfine field with increasing pressure and it completely vanishes at $\sim$ 7.5 GPa which is in contrast to the behavior in polycrystalline samples where the magnetic order vanishes at $\sim$ 20 GPa.
The different behavior of the polycrystalline samples might be due to As-vacancies.
Our results are in qualitative agreement with density functional theory calculations where a reduction of the magnetic moment with increasing pressure was found.
We found that among different samples at ambient pressure the magnetic phase transition temperature as well as the low-temperature magnetic hyperfine field decrease with increasing unit cell volume.

\end{abstract}

\maketitle

\section{introduction}

LaFeAsO as a member of the 1111 family is one of the most studied compounds of the iron-based superconductors.
It offers high superconducting transition temperatures in the case of F-substitution\cite{ja800073m,nmat2397} or multiple antiferromagnetic phases in the case of P-substitution.\cite{PhysRevB.90.064504}
Additionally LaFeAsO is theoretically approachable without the additional complication due to 3\textit{d}-4\textit{f} interaction in other rare earth 1111 compounds such as CeFeAsO, SmFeAsO, or PrFeAsO.\cite{PhysRevB.80.094524,0295-5075-84-3-37006,1367-2630-11-2-025011}
In this work we focus on the pressure dependent phase diagram on LaFeAsO single crystals.

Upon cooling LaFeAsO exhibit a structural phase transition from \textit{P}4/\textit{nmm} to \textit{Cmma} at 145 K and shows spin density wave order below 127 K.\cite{Kappenberger20189}
Measurements on polycrystalline samples have shown that the magnetic order is suppressed with increasing pressure and fully vanished for pressures of $\sim$ 20 GPa.\cite{1.4904954,JPSJ.78.123703}
On the other hand, single crystal resistivity measurements by McElroy \textit{et al}. have shown that the suppression of the magnetic order with increasing pressure is much stronger than in polycrystalline samples.\cite{PhysRevB.90.125134}
They found a nearly linear reduction of the magnetic ordering temperature to around 60 K at 6 GPa and extrapolated that the critical pressure for a full suppression of the magnetic order is 8 -- 10 GPa.
Following up on their work we investigated the microscopic magnetic phase diagram of LaFeAsO under pressure by means of synchrotron M\"ossbauer spectroscopy.
We found that LaFeAsO single crystals behave differently from polycrystalline samples and that the magnetic order is already vanished at $\sim$ 7.5 GPa.

\section{experimental details}
A LaFeAsO single crystal was investigated by means of synchrotron M\"ossbauer spectroscopy (SMS), also known as nuclear resonant forward scattering at the beamline 3ID-B of the Advanced Photon Source (APS), Argonne National Laboratory, USA.
The single crystal growth is described in detail elsewhere.\cite{Kappenberger20189}
The SMS experiments were performed in the hybrid filling operation mode of the APS in linear polarization of the beam and with a bunch separation time of 1594 ns.
This large time window allows for a high precision measurement of weak hyperfine interactions.
The single crystals were enriched with 15 \% $^{57}$Fe to ensure a sufficient count rate.
SMS spectra were recorded between 13 and 125 K and at applied pressures between 0.5 and 7.5 GPa using a special He-flow cryostat and a miniature diamond anvil cell .\cite{Bihf5283,1.4999787}
Diamond anvils with 800 $\mu$m culet size were used.
A Re gasket was preindented to 140 $\mu$m and a hole of 400 $\mu$m diameter was EDM drilled to act as a sample chamber.
Daphne oil 7575 was used as the pressure transmitting medium ensuring quasi-hydrostaticity.
The pressure was measured \textit{in situ} using an online ruby system and changed at 100 K by a gas membrane system.
The uncertainty in the pressure determination is 0.1 GPa.
The beam size was 15 $\times$ 20 $\mu$m$^2$ full width at half maximum.
CONUSS was used to analyze the SMS data.\cite{CONUSS}
For a detailed introduction into SMS the interested reader is referred to the reviews of Sturhahn.\cite{Sturhahn1998,0953-8984-16-5-009}

\section{results and discussion}
\label{sec:results}

\begin{figure*}[htbp]
	\includegraphics[width = \textwidth]{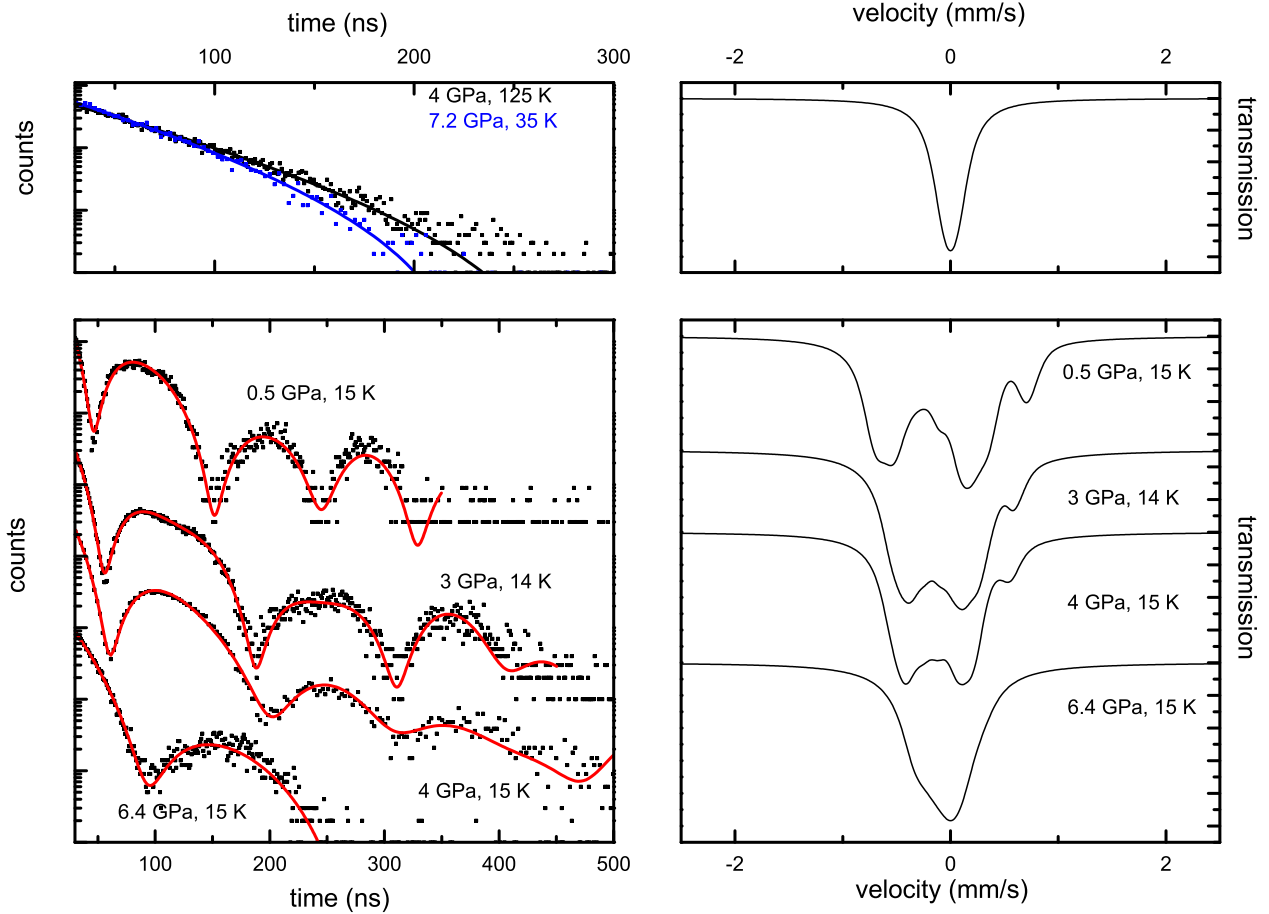}
	\caption[spectra]{Synchrotron M\"ossbauer spectroscopy spectra for representative pressures and temperatures (left column) and the calculated corresponding spectra in the energy domain (right column). Lines in the left column are fits to the data. Top row: spectra in the paramagnetic temperature regime. The absence of any quantum beats indicates no electric field gradient and no magnetic hyperfine field at the Fe nucleus. Bottom row: the quantum beating patterns show the magnetic order in the sample. The increase of the quantum beat period indicates the reduction of the magnetic hyperfine field with increasing pressure. The energy spectra in the right column were calculated using the hyperfine parameters of the corresponding time spectra.}
	\label{fig:spectra}
\end{figure*}

Synchrotron M\"ossbauer spectra for representative pressures and temperatures are shown in Fig. \ref{fig:spectra}.
No quantum beats were observed in the paramagnetic temperature regime in the investigated pressure region.
This gives an upper limit for the quadrupole splitting of $\sim$ 0.04 mm/s which corresponds to an electric field gradient at the Fe nucleus of close to zero.
This shows that the electronic environment of the Fe nucleus is nearly spherical.
Therefore the FeAs tetrahedra is uniformly compressed with increasing pressure.
Additionally this indicates a hydrostatic pressure condition.
In the magnetic phase the quadrupole splitting is $\sim$ 0.34 mm/s at 0.5 GPa and decreases to $\sim$ 0.3 mm/s and $\sim$ 0.2 mm/s at 4 GPa and 6.4 GPa, respectively.
At ambient pressure in polycrystalline samples in the magnetic phase quadrupole splittings of 0.12 to 0.3 mm/s were reported which are in fair agreement with our results.\cite{PhysRevLett.101.077005, PhysRevB.78.094517, JPSJ.78.123703}
Studies in CeFeAsO and FeSe indicate that the increase in the quadrupole splitting is a result of the magnetic ordering and that the influence of the orthorhombic distortion is negligible.\cite{PhysRevB.98.014517, Blachowski20101}

\begin{figure}[htbp]
	\includegraphics[width = \columnwidth]{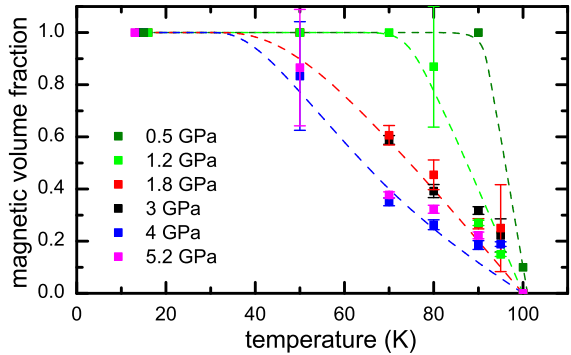}
	\caption[MVF]{Magnetic volume fraction as a function of temperature for various pressures. The dashed lines are guide to the eye only. A broadening of the magnetic phase transition area up to 5.2 GPa was observed.}
		\label{fig:MVF}
\end{figure}

\begin{figure}[htbp]
	\includegraphics[width = \columnwidth]{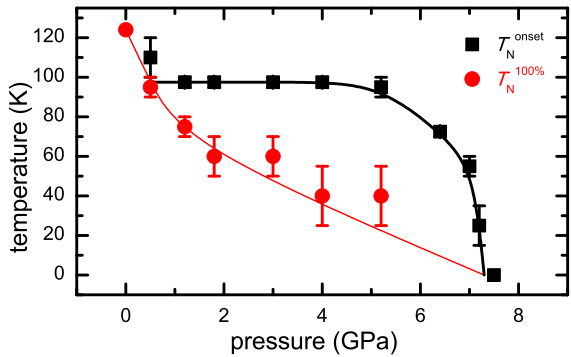}
	\caption[Tp]{Characteristic temperatures of the magnetic phase transition: \ton \,{} (black square) and \tfull \,{} (red circle) for the investigated pressure regime. \ton \,{} describes the highest temperature with a non-zero magnetic volume fraction and \tfull \,{} is the highest temperature with 100 \% MVF. For 6.4 and 7.2 GPa no magnetic volume fraction was extractable from the data and therefore only \ton \,{} is shown. Line is a guide to the eye only. Data point at ambient pressure is taken from Ref.\cite{Kappenberger20189}}
		\label{fig:Tp}
\end{figure}

In the magnetically ordered phase the SMS spectra quantum beats appear arising from the nuclear Zeeman splitting.
With increasing pressure the quantum beat period increases indicating a reduction of the magnetic hyperfine field.
The magnetic phase transition region was modeled using a paramagnetic and magnetically ordered signal fraction indicating values of the magnetic volume fraction (MVF) between zero and one.
From the temperature dependence of the MVF, which is shown in Fig. \ref{fig:MVF}, two characteristic temperatures for the magnetic phase transition, \ton \,{} and \tfull, can be extracted.
\ton \,{} describes the highest temperature with a non-zero MVF while \tfull \,{} is the highest temperature with 100 \% MVF and both are shown in Fig. \ref{fig:Tp}.
\ton \,{} remains constant within error bars at $\sim$ 95 K up to 5.2 GPa and is vanished at 7.5 GPa.

In contrast \tfull \,{} is reduced with increasing pressure to 40(15) K at 5.2 GPa.
Therefore, the magnetic phase transition region $\Delta$\textit{T} = \ton $-$ \tfull \,{} increases from 0 to $\sim$ 60 K for 0 and 5.2 GPa, respectively.
This increase in $\Delta$\textit{T} was also seen in muon spin relaxation experiments under pressure.\cite{0953-2048-25-8-084009}
An increase in $\Delta$\textit{T} is commonly attributed to a spatial distribution of \tn.\cite{PhysRevB.92.134511, PhysRevB.98.014517, PhysRevB.89.144511,PhysRevB.87.174519}
A possible cause could be an increased strain from lattice misfit with increasing pressure.\cite{PhysRevB.82.144507}
For 6.4 and 7.2 GPa an extraction of the MVF was not possible due to small magnetic hyperfine fields.
Thus the MVF was set to one during the analysis of those pressures in the magnetic phase transition region.
However, this does not influence the analysis of the low-temperature data.

\begin{figure}[htbp]
	\includegraphics[width = \columnwidth]{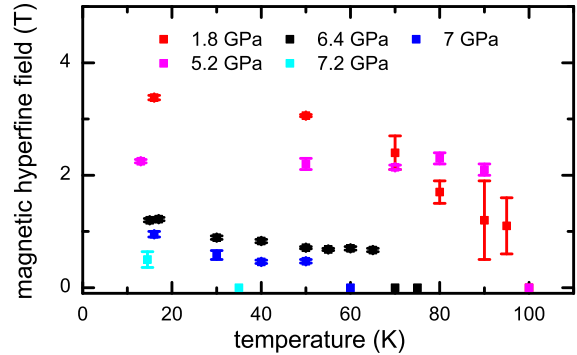}
		\caption[BT]{Magnetic hyperfine field as a function of temperature for various pressures (some are omitted for the sake of clarity). A reduction in the low-temperature magnetic hyperfine field was observed.}
		\label{fig:BT}
\end{figure}

The magnetic hyperfine field as a function of temperature for representative pressure values is shown in Fig. \ref{fig:BT}.
A reduction of the magnetic hyperfine field with increasing pressure was observed.

\begin{figure}[htbp]
	\includegraphics[width = \columnwidth]{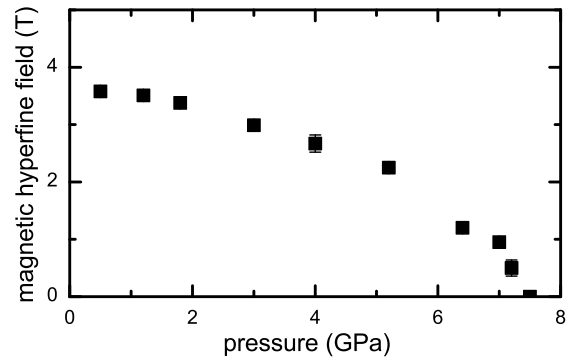}
		\caption[Bp]{Low-temperature magnetic hyperfine field for all investigated pressures. The magnetic hyperfine field is continuously reduced to zero at 7.5 GPa.}
		\label{fig:Bpd}
\end{figure}

The low-temperature magnetic hyperfine field as a function of pressure is shown in Fig. \ref{fig:Bpd}.
The low-temperature magnetic hyperfine field is continuously reduced to zero at 7.5 GPa.
For this pressure an Fe-As distance of $\sim$ 2.37 \AA \,{} can be extrapolated from reported room temperature data.\cite{Kobayashi2016}

 
It was shown that the Fe magnetic moment and thus the magnetic hyperfine field are related to the Fe-As distance which controls the Fe 3\textit{d}-As 4\textit{p} hybridization strength.\cite{PhysRevB.67.155421}
If the hybridization is strong enough the Fe magnetic moment is quenched.\cite{PhysRevB.78.024521, PhysRevLett.101.126401}
Therefore the continuous reduction of the magnetic hyperfine field to zero with decreasing Fe-As distance supports the picture that the \textit{dp} hybridization strength controls the value of the iron magnetic moment.
Theoretical calculations suggested that the critical Fe-As distance where the Fe magnetic moment vanishes is 2.36 \AA.\cite{PhysRevB.67.155421}
In our study the critical Fe-As distance is estimated to be $\sim$ 2.37 \AA \,{} which is in good agreement with the calculations.
Additionally our results are in qualitative agreement with density functional theory calculations where a reduction of the magnetic moment with increasing pressure was found.\cite{PhysRevB.79.024509}
%


Extrapolating the obtained pressure dependence of the magnetic hyperfine field to zero pressure results in 3.7(1) T which is 1.3 $-$ 1.6 T smaller than for LaFeAsO polycrystalline samples.\cite{PhysRevLett.101.077005, PhysRevB.78.094517,1367-2630-11-2-025011}
Additionally the magnetic phase transition temperature \tn \,{} determined by electrical resistivity, magnetic susceptibility, and specific-heat measurements on single crystals\cite{Kappenberger20189, PhysRevB.88.134513, PhysRevB.90.125134, PhysRevB.86.134511} has values between 117 and 127 K which are up to 36 K smaller than in polycrystalline samples.\cite{PhysRevLett.101.077005, JPSJ.78.123703,PhysRevB.78.094517}
Both \tn \,{} and the low-temperature magnetic hyperfine field \textit{B} for single and polycrystalline samples are summarized in Tab. \ref{tab:LTB1111}.

\begin{table}[htbp]
\caption{Low-temperature magnetic hyperfine field \textit{B} for LaFeAsO as well as the magnetic phase transition temperatures \tn \,{} at ambient pressure (if not stated otherwise).}
	\centering
		\begin{tabular}{cccc}
		\hline\hline
		\textit{B} / T 	&	\tn / K	&\\
		3.7(1)	&				127			 & single crystal\cite{Kappenberger20189}\\
		4.86(5)	&			138	 & polycrystal\cite{PhysRevLett.101.077005}	\\
		5.1			&			139	& polycrystal\cite{RAFFIUS1993135}\\
		5.19(1)	&			153	 & polycrystal \cite{PhysRevB.78.094517}\\
		5.3	&					145(5)		& polycrystal, 0.1 MPa\cite{JPSJ.78.123703}\\	
		5.5	&		&	polycrystal, 4 GPa\cite{1.4904954}\\
\hline\hline
\end{tabular}
	\label{tab:LTB1111}
\end{table}

Samples with smaller \tn \,{} also show a smaller \textit{B}.
This can be qualitatively understood in the framework of Landau theory where the magnetic order parameter \textit{M} is proportional to \tn \,{} with $M \propto \sqrt{T_{\text{N}}}$.
Therefore a reduction of \tn \,{} results in a reduction of \textit{M} and thus of \textit{B}.

\begin{figure}[htbp]
	\includegraphics[width = \columnwidth]{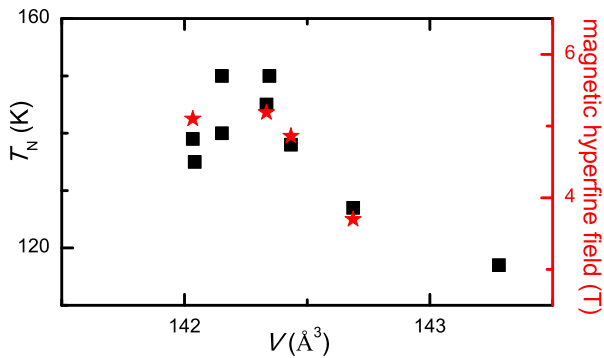}
		\caption[TN-B-V]{\tn \,{} (black square) and low-temperature magnetic hyperfine field \textit{B} (red star) as a function of the room temperature unit cell volume \textit{V}. Data taken from Refs.\cite{Kappenberger20189,PhysRevB.88.134513,PhysRevB.78.094517,PhysRevB.86.134511,RAFFIUS1993135,JPSJ.77.113712,PhysRevB.82.184521,0953-2048-21-12-125028,nmat2397} By increasing the unit cell volume both \tn \,{} and \textit{B} decrease.}
		\label{fig:TNBV}
\end{figure}

Additionally different samples not only deviate in \tn \,{} and \textit{B} but also in the crystallographic parameters.
This is shown in Fig. \ref{fig:TNBV}.
The crystallographic \textit{a} axis varies between 4.0367 \AA  \,{} and 4.0308 \AA  \,{} and thus is changed by $<$0.15 \%.\cite{Kappenberger20189,PhysRevB.88.134513,PhysRevB.78.094517,PhysRevB.86.134511,RAFFIUS1993135,JPSJ.77.113712,PhysRevB.82.184521,0953-2048-21-12-125028,nmat2397}
In contrast the crystallographic \textit{c} axis varies between 8.793 \AA \,{} and 8.7364
 \AA  \,{} and thus is changed by up to 0.65 \%.\cite{Kappenberger20189,PhysRevB.88.134513,PhysRevB.78.094517,PhysRevB.86.134511,RAFFIUS1993135,JPSJ.77.113712,PhysRevB.82.184521,0953-2048-21-12-125028,nmat2397}
By increasing the unit cell volume and in particular the crystallographic \textit{c} axis both \tn \,{} and \textit{B} decrease.
Theoretical calculations suggest that the interlayer coupling is weak but important to stabilize the magnetic order.\cite{ANDP201000149,PhysRevB.78.020501,0295-5075-88-1-17010}
By increasing the crystallographic \textit{c} axis and thus the interlayer distance the interlayer coupling may decrease resulting in a weakened magnetic order.
This theoretical picture is supported by the reduction of \tn \,{} and \textit{B} with increasing \textit{c}.
However, the origin of the discrepancy in the crystallographic parameters among different samples is unknown.


Resistivity measurements on polycrystalline samples show a linear reduction of the magnetic phase transition temperature between 0 and 2 GPa followed by an upturn.\cite{JPSJ.77.113712}
In contrast resistivity measurements on a single crystal show a linear reduction of \tn \,{} to 60 K at 6 GPa.\cite{PhysRevB.90.125134}
Energy-domain M\"ossbauer spectroscopy measurements of polycrystalline samples under pressure have shown 100 \% magnetic volume fraction at 8 GPa and 8 K.\cite{JPSJ.78.123703}
At pressures  $>$ 8 GPa the magnetic volume fraction decreases with increasing pressure until a pure paramagnetic signal is observed at 24 GPa and 8 K.\cite{JPSJ.78.123703}
The pressure dependence of the reported magnetic hyperfine field follows the pressure dependence of the magnetic phase transition temperature determined by resistivity measurements in polycrystalline samples.\cite{JPSJ.77.113712,JPSJ.78.123703}
It shows a linear reduction between 0 and $\sim$ 2 GPa followed by a plateau up to $\sim$ 20 GPa and a subsequent reduction to zero.\cite{JPSJ.78.123703}
In contrast, the magnetic hyperfine field obtained from SMS data on a single crystal shows a continuous reduction to zero at 7.5 GPa.
The combination of the resistivity and M\"ossbauer experiments indicate that the single- and polycrystalline samples behave qualitatively similar at pressures below 2 GPa but differ at higher pressures.

The large difference between single and polycrystals may be due to the granular and inhomogeneous nature and O-deficiency at the grain boundaries of the latter. 
It was pointed out by McElroy \textit{et al}. that in polycrystalline samples the reduction in the resistivity at lower temperatures is very broad and it goes to zero only at 12 GPa which is maybe caused by tiny amounts of O-deficient sample volumes.\cite{PhysRevB.90.125134,JPSJ.77.113712}
M\"ossbauer measurements by Nowik \textit{et al}. on O-deficient LaFeAsO have shown that Fe in the vicinity of an O vacancy has a small magnetic hyperfine field of $\sim$ 0.8 T but a huge quadrupole splitting of $\sim$ $-$0.86 mm/s at low temperatures.\cite{0953-8984-20-29-292201}
In the M\"ossbauer experiments under pressure of the polycrystalline samples no O-deficiency was detected.\cite{JPSJ.78.123703}
Taking into account the volume resolution of the method ($\sim$ 1 \%) it is in agreement with possible tiny amounts of O-vacancies causing filamentary superconductivity.\cite{JPSJ.78.123703, PhysRevB.90.125134}
Additionally the magnetic hyperfine field of $\sim$ 0.8 T is too small to account for the oberserved plateau at $\sim$ 3 T and therefore a significant O-deficiency can be ruled out.\cite{JPSJ.78.123703}

Another possible explanation are As-vacancies in the polycrystalline samples.
It was shown in LaFeAs$_{1-x}$O$_{0.9}$F$_{0.1}$ that As-vacancies act as magnetic defects with a magnetic moment of 0.8 $\mu_{\mathrm{B}}$/Fe due to a spin polarization of the Fe 3\textit{d} electrons if the Fe 3\textit{d}-As 4\textit{p} hybridization is sufficiently strong enough.\cite{PhysRevB.84.134516}
This might lead to the situation that with increasing pressure the Fe 3\textit{d}-As 4\textit{p} hybridization will reduce the magnetic moment but the As-vacancies will enhance the spin polarization of the Fe 3\textit{d} electrons and thus stabilize the magnetic order.
To support or falsify this possibility further investigations are needed.


%
%

\section{summary and conclusion}
In summary we conducted synchrotron M\"ossbauer experiments at pressures up to 7.5 GPa and at temperatures between 13 and 125 K and provide a microscopic phase diagram of LaFeAsO single crystals under pressure.
At the magnetic phase transition an increase in the quadrupole splitting was observed which is most likely of magnetic origin.
The magnetic hyperfine field is continuously suppressed to zero at $\sim$ 7.5 GPa which corresponds to a Fe-As distance of $\sim$ 2.37 \AA.
Our results indicate that single and polycrystalline samples behave qualitatively similar up to 2 GPa but differ at higher pressures.
Possible cause in polycrystalline samples could be due to their granular and inhomogeneous nature and O-deficiency at the grain boundaries.  
Another possibility in polycrystalline samples could be As-vacancies acting as magnetic centers as shown in Ref.\cite{PhysRevB.84.134516}.
We found that among different samples the magnetic phase transition temperature as well as the low-temperature magnetic hyperfine field decrease with increasing unit cell volume which might explain the difference in the observed quantities.
Interestingly the behavior of both LaFeAsO single crystals\cite{PhysRevB.88.134513, PhysRevB.90.125134, PhysRevLett.109.247001, PhysRevB.86.134511, PhysRevB.97.180405} and polycrystalline samples\cite{PhysRevLett.101.077005, JPSJ.78.123703, 1.4904954, JPSJ.77.113712} are consistent within each other.

\acknowledgments
Part of this work was funded by the Deutsche Forschungsgemeinschaft (DFG, German Research Foundation) -- MA 7362/1-1, WU 595/3-3, BU 887/15-1, and the research training group GRK-1621.
This research used resources of the Advanced Photon Source, a U.S. Department of Energy (DOE) Office of Science User Facility operated for the DOE Office of Science by Argonne National Laboratory under Contract No. DE-AC02-06CH11357.
Support from COMPRES under NSF Cooperative Agreement EAR-1606856 is acknowledged for partial support of W. Bi.
\bibliography{LaFeAsO}

\begin{thebibliography}{42}%
\makeatletter
\providecommand \@ifxundefined [1]{%
 \@ifx{#1\undefined}
}%
\providecommand \@ifnum [1]{%
 \ifnum #1\expandafter \@firstoftwo
 \else \expandafter \@secondoftwo
 \fi
}%
\providecommand \@ifx [1]{%
 \ifx #1\expandafter \@firstoftwo
 \else \expandafter \@secondoftwo
 \fi
}%
\providecommand \natexlab [1]{#1}%
\providecommand \enquote  [1]{``#1''}%
\providecommand \bibnamefont  [1]{#1}%
\providecommand \bibfnamefont [1]{#1}%
\providecommand \citenamefont [1]{#1}%
\providecommand \href@noop [0]{\@secondoftwo}%
\providecommand \href [0]{\begingroup \@sanitize@url \@href}%
\providecommand \@href[1]{\@@startlink{#1}\@@href}%
\providecommand \@@href[1]{\endgroup#1\@@endlink}%
\providecommand \@sanitize@url [0]{\catcode `\\12\catcode `\$12\catcode
  `\&12\catcode `\#12\catcode `\^12\catcode `\_12\catcode `\%12\relax}%
\providecommand \@@startlink[1]{}%
\providecommand \@@endlink[0]{}%
\providecommand \url  [0]{\begingroup\@sanitize@url \@url }%
\providecommand \@url [1]{\endgroup\@href {#1}{\urlprefix }}%
\providecommand \urlprefix  [0]{URL }%
\providecommand \Eprint [0]{\href }%
\providecommand \doibase [0]{http://dx.doi.org/}%
\providecommand \selectlanguage [0]{\@gobble}%
\providecommand \bibinfo  [0]{\@secondoftwo}%
\providecommand \bibfield  [0]{\@secondoftwo}%
\providecommand \translation [1]{[#1]}%
\providecommand \BibitemOpen [0]{}%
\providecommand \bibitemStop [0]{}%
\providecommand \bibitemNoStop [0]{.\EOS\space}%
\providecommand \EOS [0]{\spacefactor3000\relax}%
\providecommand \BibitemShut  [1]{\csname bibitem#1\endcsname}%
\let\auto@bib@innerbib\@empty
\bibitem [{\citenamefont {Kamihara}\ \emph {et~al.}(2008)\citenamefont
  {Kamihara}, \citenamefont {Watanabe}, \citenamefont {Hirano},\ and\
  \citenamefont {Hosono}}]{ja800073m}%
  \BibitemOpen
  \bibfield  {author} {\bibinfo {author} {\bibfnamefont {Y.}~\bibnamefont
  {Kamihara}}, \bibinfo {author} {\bibfnamefont {T.}~\bibnamefont {Watanabe}},
  \bibinfo {author} {\bibfnamefont {M.}~\bibnamefont {Hirano}}, \ and\ \bibinfo
  {author} {\bibfnamefont {H.}~\bibnamefont {Hosono}},\ }\href {\doibase
  10.1021/ja800073m} {\bibfield  {journal} {\bibinfo  {journal} {J. Am. Chem.
  Soc.}\ }\textbf {\bibinfo {volume} {130}},\ \bibinfo {pages} {3296} (\bibinfo
  {year} {2008})}\BibitemShut {NoStop}%
\bibitem [{\citenamefont {Luetkens}\ \emph {et~al.}(2009)\citenamefont
  {Luetkens}, \citenamefont {Klauss}, \citenamefont {Kraken}, \citenamefont
  {Litterst}, \citenamefont {Dellmann}, \citenamefont {Klingeler},
  \citenamefont {Hess}, \citenamefont {Khasanov}, \citenamefont {Amato},
  \citenamefont {Baines}, \citenamefont {Kosmala}, \citenamefont {Schumann},
  \citenamefont {Braden}, \citenamefont {Hamann-Borrero}, \citenamefont {Leps},
  \citenamefont {Kondrat}, \citenamefont {Behr}, \citenamefont {Werner},\ and\
  \citenamefont {Buchner}}]{nmat2397}%
  \BibitemOpen
  \bibfield  {author} {\bibinfo {author} {\bibfnamefont {H.}~\bibnamefont
  {Luetkens}}, \bibinfo {author} {\bibfnamefont {H.-H.}\ \bibnamefont
  {Klauss}}, \bibinfo {author} {\bibfnamefont {M.}~\bibnamefont {Kraken}},
  \bibinfo {author} {\bibfnamefont {F.~J.}\ \bibnamefont {Litterst}}, \bibinfo
  {author} {\bibfnamefont {T.}~\bibnamefont {Dellmann}}, \bibinfo {author}
  {\bibfnamefont {R.}~\bibnamefont {Klingeler}}, \bibinfo {author}
  {\bibfnamefont {C.}~\bibnamefont {Hess}}, \bibinfo {author} {\bibfnamefont
  {R.}~\bibnamefont {Khasanov}}, \bibinfo {author} {\bibfnamefont
  {A.}~\bibnamefont {Amato}}, \bibinfo {author} {\bibfnamefont
  {C.}~\bibnamefont {Baines}}, \bibinfo {author} {\bibfnamefont
  {M.}~\bibnamefont {Kosmala}}, \bibinfo {author} {\bibfnamefont {O.~J.}\
  \bibnamefont {Schumann}}, \bibinfo {author} {\bibfnamefont {M.}~\bibnamefont
  {Braden}}, \bibinfo {author} {\bibfnamefont {J.}~\bibnamefont
  {Hamann-Borrero}}, \bibinfo {author} {\bibfnamefont {N.}~\bibnamefont
  {Leps}}, \bibinfo {author} {\bibfnamefont {A.}~\bibnamefont {Kondrat}},
  \bibinfo {author} {\bibfnamefont {G.}~\bibnamefont {Behr}}, \bibinfo {author}
  {\bibfnamefont {J.}~\bibnamefont {Werner}}, \ and\ \bibinfo {author}
  {\bibfnamefont {B.}~\bibnamefont {Buchner}},\ }\href {\doibase
  10.1038/nmat2397} {\bibfield  {journal} {\bibinfo  {journal} {Nat. Mater.}\
  }\textbf {\bibinfo {volume} {8}},\ \bibinfo {pages} {305} (\bibinfo {year}
  {2009})}\BibitemShut {NoStop}%
\bibitem [{\citenamefont {Lai}\ \emph {et~al.}(2014)\citenamefont {Lai},
  \citenamefont {Takemori}, \citenamefont {Miyasaka}, \citenamefont {Engetsu},
  \citenamefont {Mukuda},\ and\ \citenamefont {Tajima}}]{PhysRevB.90.064504}%
  \BibitemOpen
  \bibfield  {author} {\bibinfo {author} {\bibfnamefont {K.~T.}\ \bibnamefont
  {Lai}}, \bibinfo {author} {\bibfnamefont {A.}~\bibnamefont {Takemori}},
  \bibinfo {author} {\bibfnamefont {S.}~\bibnamefont {Miyasaka}}, \bibinfo
  {author} {\bibfnamefont {F.}~\bibnamefont {Engetsu}}, \bibinfo {author}
  {\bibfnamefont {H.}~\bibnamefont {Mukuda}}, \ and\ \bibinfo {author}
  {\bibfnamefont {S.}~\bibnamefont {Tajima}},\ }\href {\doibase
  10.1103/PhysRevB.90.064504} {\bibfield  {journal} {\bibinfo  {journal} {Phys.
  Rev. B}\ }\textbf {\bibinfo {volume} {90}},\ \bibinfo {pages} {064504}
  (\bibinfo {year} {2014})}\BibitemShut {NoStop}%
\bibitem [{\citenamefont {Maeter}\ \emph {et~al.}(2009)\citenamefont {Maeter},
  \citenamefont {Luetkens}, \citenamefont {Pashkevich}, \citenamefont
  {Kwadrin}, \citenamefont {Khasanov}, \citenamefont {Amato}, \citenamefont
  {Gusev}, \citenamefont {Lamonova}, \citenamefont {Chervinskii}, \citenamefont
  {Klingeler}, \citenamefont {Hess}, \citenamefont {Behr}, \citenamefont
  {B\"uchner},\ and\ \citenamefont {Klauss}}]{PhysRevB.80.094524}%
  \BibitemOpen
  \bibfield  {author} {\bibinfo {author} {\bibfnamefont {H.}~\bibnamefont
  {Maeter}}, \bibinfo {author} {\bibfnamefont {H.}~\bibnamefont {Luetkens}},
  \bibinfo {author} {\bibfnamefont {Y.~G.}\ \bibnamefont {Pashkevich}},
  \bibinfo {author} {\bibfnamefont {A.}~\bibnamefont {Kwadrin}}, \bibinfo
  {author} {\bibfnamefont {R.}~\bibnamefont {Khasanov}}, \bibinfo {author}
  {\bibfnamefont {A.}~\bibnamefont {Amato}}, \bibinfo {author} {\bibfnamefont
  {A.~A.}\ \bibnamefont {Gusev}}, \bibinfo {author} {\bibfnamefont {K.~V.}\
  \bibnamefont {Lamonova}}, \bibinfo {author} {\bibfnamefont {D.~A.}\
  \bibnamefont {Chervinskii}}, \bibinfo {author} {\bibfnamefont
  {R.}~\bibnamefont {Klingeler}}, \bibinfo {author} {\bibfnamefont
  {C.}~\bibnamefont {Hess}}, \bibinfo {author} {\bibfnamefont {G.}~\bibnamefont
  {Behr}}, \bibinfo {author} {\bibfnamefont {B.}~\bibnamefont {B\"uchner}}, \
  and\ \bibinfo {author} {\bibfnamefont {H.-H.}\ \bibnamefont {Klauss}},\
  }\href {\doibase 10.1103/PhysRevB.80.094524} {\bibfield  {journal} {\bibinfo
  {journal} {Phys. Rev. B}\ }\textbf {\bibinfo {volume} {80}},\ \bibinfo
  {pages} {094524} (\bibinfo {year} {2009})}\BibitemShut {NoStop}%
\bibitem [{\citenamefont {Pourovskii}\ \emph {et~al.}(2008)\citenamefont
  {Pourovskii}, \citenamefont {Vildosola}, \citenamefont {Biermann},\ and\
  \citenamefont {Georges}}]{0295-5075-84-3-37006}%
  \BibitemOpen
  \bibfield  {author} {\bibinfo {author} {\bibfnamefont {L.}~\bibnamefont
  {Pourovskii}}, \bibinfo {author} {\bibfnamefont {V.}~\bibnamefont
  {Vildosola}}, \bibinfo {author} {\bibfnamefont {S.}~\bibnamefont {Biermann}},
  \ and\ \bibinfo {author} {\bibfnamefont {A.}~\bibnamefont {Georges}},\ }\href
  {http://stacks.iop.org/0295-5075/84/i=3/a=37006} {\bibfield  {journal}
  {\bibinfo  {journal} {Europhys. Lett.}\ }\textbf {\bibinfo {volume} {84}},\
  \bibinfo {pages} {37006} (\bibinfo {year} {2008})}\BibitemShut {NoStop}%
\bibitem [{\citenamefont {McGuire}\ \emph {et~al.}(2009)\citenamefont
  {McGuire}, \citenamefont {Hermann}, \citenamefont {Sefat}, \citenamefont
  {Sales}, \citenamefont {Jin}, \citenamefont {Mandrus}, \citenamefont
  {Grandjean},\ and\ \citenamefont {Long}}]{1367-2630-11-2-025011}%
  \BibitemOpen
  \bibfield  {author} {\bibinfo {author} {\bibfnamefont {M.~A.}\ \bibnamefont
  {McGuire}}, \bibinfo {author} {\bibfnamefont {R.~P.}\ \bibnamefont
  {Hermann}}, \bibinfo {author} {\bibfnamefont {A.~S.}\ \bibnamefont {Sefat}},
  \bibinfo {author} {\bibfnamefont {B.~C.}\ \bibnamefont {Sales}}, \bibinfo
  {author} {\bibfnamefont {R.}~\bibnamefont {Jin}}, \bibinfo {author}
  {\bibfnamefont {D.}~\bibnamefont {Mandrus}}, \bibinfo {author} {\bibfnamefont
  {F.}~\bibnamefont {Grandjean}}, \ and\ \bibinfo {author} {\bibfnamefont
  {G.~J.}\ \bibnamefont {Long}},\ }\href
  {http://stacks.iop.org/1367-2630/11/i=2/a=025011} {\bibfield  {journal}
  {\bibinfo  {journal} {N. J. Phys.}\ }\textbf {\bibinfo {volume} {11}},\
  \bibinfo {pages} {025011} (\bibinfo {year} {2009})}\BibitemShut {NoStop}%
\bibitem [{\citenamefont {Kappenberger}\ \emph {et~al.}(2018)\citenamefont
  {Kappenberger}, \citenamefont {Aswartham}, \citenamefont {Scaravaggi},
  \citenamefont {Blum}, \citenamefont {Sturza}, \citenamefont {Wolter},
  \citenamefont {Wurmehl},\ and\ \citenamefont {Büchner}}]{Kappenberger20189}%
  \BibitemOpen
  \bibfield  {author} {\bibinfo {author} {\bibfnamefont {R.}~\bibnamefont
  {Kappenberger}}, \bibinfo {author} {\bibfnamefont {S.}~\bibnamefont
  {Aswartham}}, \bibinfo {author} {\bibfnamefont {F.}~\bibnamefont
  {Scaravaggi}}, \bibinfo {author} {\bibfnamefont {C.~G.}\ \bibnamefont
  {Blum}}, \bibinfo {author} {\bibfnamefont {M.~I.}\ \bibnamefont {Sturza}},
  \bibinfo {author} {\bibfnamefont {A.~U.}\ \bibnamefont {Wolter}}, \bibinfo
  {author} {\bibfnamefont {S.}~\bibnamefont {Wurmehl}}, \ and\ \bibinfo
  {author} {\bibfnamefont {B.}~\bibnamefont {Büchner}},\ }\href {\doibase
  10.1016/j.jcrysgro.2017.11.006} {\bibfield  {journal} {\bibinfo  {journal}
  {J. Cryst. Growth}\ }\textbf {\bibinfo {volume} {483}},\ \bibinfo {pages} {9
  } (\bibinfo {year} {2018})}\BibitemShut {NoStop}%
\bibitem [{\citenamefont {Kumar}\ \emph {et~al.}(2014)\citenamefont {Kumar},
  \citenamefont {Hamlin}, \citenamefont {Maple}, \citenamefont {Zhang},
  \citenamefont {Chen}, \citenamefont {Baker}, \citenamefont {Cornelius},
  \citenamefont {Zhao}, \citenamefont {Xiao}, \citenamefont {Sinogeikin},\ and\
  \citenamefont {Chow}}]{1.4904954}%
  \BibitemOpen
  \bibfield  {author} {\bibinfo {author} {\bibfnamefont {R.~S.}\ \bibnamefont
  {Kumar}}, \bibinfo {author} {\bibfnamefont {J.~J.}\ \bibnamefont {Hamlin}},
  \bibinfo {author} {\bibfnamefont {M.~B.}\ \bibnamefont {Maple}}, \bibinfo
  {author} {\bibfnamefont {Y.}~\bibnamefont {Zhang}}, \bibinfo {author}
  {\bibfnamefont {C.}~\bibnamefont {Chen}}, \bibinfo {author} {\bibfnamefont
  {J.}~\bibnamefont {Baker}}, \bibinfo {author} {\bibfnamefont {A.~L.}\
  \bibnamefont {Cornelius}}, \bibinfo {author} {\bibfnamefont {Y.}~\bibnamefont
  {Zhao}}, \bibinfo {author} {\bibfnamefont {Y.}~\bibnamefont {Xiao}}, \bibinfo
  {author} {\bibfnamefont {S.}~\bibnamefont {Sinogeikin}}, \ and\ \bibinfo
  {author} {\bibfnamefont {P.}~\bibnamefont {Chow}},\ }\href {\doibase
  10.1063/1.4904954} {\bibfield  {journal} {\bibinfo  {journal} {Appl. Phys.
  Lett.}\ }\textbf {\bibinfo {volume} {105}},\ \bibinfo {pages} {251902}
  (\bibinfo {year} {2014})}\BibitemShut {NoStop}%
\bibitem [{\citenamefont {Kawakami}\ \emph {et~al.}(2009)\citenamefont
  {Kawakami}, \citenamefont {Kamatani}, \citenamefont {Okada}, \citenamefont
  {Takahashi}, \citenamefont {Nasu}, \citenamefont {Kamihara}, \citenamefont
  {Hirano},\ and\ \citenamefont {Hosono}}]{JPSJ.78.123703}%
  \BibitemOpen
  \bibfield  {author} {\bibinfo {author} {\bibfnamefont {T.}~\bibnamefont
  {Kawakami}}, \bibinfo {author} {\bibfnamefont {T.}~\bibnamefont {Kamatani}},
  \bibinfo {author} {\bibfnamefont {H.}~\bibnamefont {Okada}}, \bibinfo
  {author} {\bibfnamefont {H.}~\bibnamefont {Takahashi}}, \bibinfo {author}
  {\bibfnamefont {S.}~\bibnamefont {Nasu}}, \bibinfo {author} {\bibfnamefont
  {Y.}~\bibnamefont {Kamihara}}, \bibinfo {author} {\bibfnamefont
  {M.}~\bibnamefont {Hirano}}, \ and\ \bibinfo {author} {\bibfnamefont
  {H.}~\bibnamefont {Hosono}},\ }\href {\doibase 10.1143/JPSJ.78.123703}
  {\bibfield  {journal} {\bibinfo  {journal} {J. Phys. Soc. Jpn.}\ }\textbf
  {\bibinfo {volume} {78}},\ \bibinfo {pages} {123703} (\bibinfo {year}
  {2009})}\BibitemShut {NoStop}%
\bibitem [{\citenamefont {McElroy}\ \emph {et~al.}(2014)\citenamefont
  {McElroy}, \citenamefont {Hamlin}, \citenamefont {White}, \citenamefont
  {Weir}, \citenamefont {Vohra},\ and\ \citenamefont
  {Maple}}]{PhysRevB.90.125134}%
  \BibitemOpen
  \bibfield  {author} {\bibinfo {author} {\bibfnamefont {C.~A.}\ \bibnamefont
  {McElroy}}, \bibinfo {author} {\bibfnamefont {J.~J.}\ \bibnamefont {Hamlin}},
  \bibinfo {author} {\bibfnamefont {B.~D.}\ \bibnamefont {White}}, \bibinfo
  {author} {\bibfnamefont {S.~T.}\ \bibnamefont {Weir}}, \bibinfo {author}
  {\bibfnamefont {Y.~K.}\ \bibnamefont {Vohra}}, \ and\ \bibinfo {author}
  {\bibfnamefont {M.~B.}\ \bibnamefont {Maple}},\ }\href {\doibase
  10.1103/PhysRevB.90.125134} {\bibfield  {journal} {\bibinfo  {journal} {Phys.
  Rev. B}\ }\textbf {\bibinfo {volume} {90}},\ \bibinfo {pages} {125134}
  (\bibinfo {year} {2014})}\BibitemShut {NoStop}%
\bibitem [{\citenamefont {Bi}\ \emph {et~al.}(2015)\citenamefont {Bi},
  \citenamefont {Zhao}, \citenamefont {Lin}, \citenamefont {Jia}, \citenamefont
  {Hu}, \citenamefont {Jin}, \citenamefont {Ferry}, \citenamefont {Yang},
  \citenamefont {Struzhkin},\ and\ \citenamefont {Alp}}]{Bihf5283}%
  \BibitemOpen
  \bibfield  {author} {\bibinfo {author} {\bibfnamefont {W.}~\bibnamefont
  {Bi}}, \bibinfo {author} {\bibfnamefont {J.}~\bibnamefont {Zhao}}, \bibinfo
  {author} {\bibfnamefont {J.-F.}\ \bibnamefont {Lin}}, \bibinfo {author}
  {\bibfnamefont {Q.}~\bibnamefont {Jia}}, \bibinfo {author} {\bibfnamefont
  {M.~Y.}\ \bibnamefont {Hu}}, \bibinfo {author} {\bibfnamefont
  {C.}~\bibnamefont {Jin}}, \bibinfo {author} {\bibfnamefont {R.}~\bibnamefont
  {Ferry}}, \bibinfo {author} {\bibfnamefont {W.}~\bibnamefont {Yang}},
  \bibinfo {author} {\bibfnamefont {V.}~\bibnamefont {Struzhkin}}, \ and\
  \bibinfo {author} {\bibfnamefont {E.~E.}\ \bibnamefont {Alp}},\ }\href
  {\doibase 10.1107/S1600577515003586} {\bibfield  {journal} {\bibinfo
  {journal} {J. Synch. Radiat.}\ }\textbf {\bibinfo {volume} {22}},\ \bibinfo
  {pages} {760} (\bibinfo {year} {2015})}\BibitemShut {NoStop}%
\bibitem [{\citenamefont {Zhao}\ \emph {et~al.}(2017)\citenamefont {Zhao},
  \citenamefont {Bi}, \citenamefont {Sinogeikin}, \citenamefont {Hu},
  \citenamefont {Alp}, \citenamefont {Wang}, \citenamefont {Jin},\ and\
  \citenamefont {Lin}}]{1.4999787}%
  \BibitemOpen
  \bibfield  {author} {\bibinfo {author} {\bibfnamefont {J.~Y.}\ \bibnamefont
  {Zhao}}, \bibinfo {author} {\bibfnamefont {W.}~\bibnamefont {Bi}}, \bibinfo
  {author} {\bibfnamefont {S.}~\bibnamefont {Sinogeikin}}, \bibinfo {author}
  {\bibfnamefont {M.~Y.}\ \bibnamefont {Hu}}, \bibinfo {author} {\bibfnamefont
  {E.~E.}\ \bibnamefont {Alp}}, \bibinfo {author} {\bibfnamefont {X.~C.}\
  \bibnamefont {Wang}}, \bibinfo {author} {\bibfnamefont {C.~Q.}\ \bibnamefont
  {Jin}}, \ and\ \bibinfo {author} {\bibfnamefont {J.~F.}\ \bibnamefont
  {Lin}},\ }\href {\doibase 10.1063/1.4999787} {\bibfield  {journal} {\bibinfo
  {journal} {Rev. Sci. Instrum.}\ }\textbf {\bibinfo {volume} {88}},\ \bibinfo
  {pages} {125109} (\bibinfo {year} {2017})}\BibitemShut {NoStop}%
\bibitem [{\citenamefont {Sturhahn}(2000)}]{CONUSS}%
  \BibitemOpen
  \bibfield  {author} {\bibinfo {author} {\bibfnamefont {W.}~\bibnamefont
  {Sturhahn}},\ }\href {\doibase 10.1023/A:1012681503686} {\bibfield  {journal}
  {\bibinfo  {journal} {Hyperfine Interact.}\ }\textbf {\bibinfo {volume}
  {125}},\ \bibinfo {pages} {149} (\bibinfo {year} {2000})}\BibitemShut
  {NoStop}%
\bibitem [{\citenamefont {Sturhahn}\ \emph {et~al.}(1998)\citenamefont
  {Sturhahn}, \citenamefont {Alp}, \citenamefont {Toellner}, \citenamefont
  {Hession}, \citenamefont {Hu},\ and\ \citenamefont {Sutter}}]{Sturhahn1998}%
  \BibitemOpen
  \bibfield  {author} {\bibinfo {author} {\bibfnamefont {W.}~\bibnamefont
  {Sturhahn}}, \bibinfo {author} {\bibfnamefont {E.}~\bibnamefont {Alp}},
  \bibinfo {author} {\bibfnamefont {T.}~\bibnamefont {Toellner}}, \bibinfo
  {author} {\bibfnamefont {P.}~\bibnamefont {Hession}}, \bibinfo {author}
  {\bibfnamefont {M.}~\bibnamefont {Hu}}, \ and\ \bibinfo {author}
  {\bibfnamefont {J.}~\bibnamefont {Sutter}},\ }\href {\doibase
  10.1023/A:1012607212694} {\bibfield  {journal} {\bibinfo  {journal}
  {Hyperfine Interact.}\ }\textbf {\bibinfo {volume} {113}},\ \bibinfo {pages}
  {47} (\bibinfo {year} {1998})}\BibitemShut {NoStop}%
\bibitem [{\citenamefont {Sturhahn}(2004)}]{0953-8984-16-5-009}%
  \BibitemOpen
  \bibfield  {author} {\bibinfo {author} {\bibfnamefont {W.}~\bibnamefont
  {Sturhahn}},\ }\href {http://stacks.iop.org/0953-8984/16/i=5/a=009}
  {\bibfield  {journal} {\bibinfo  {journal} {J. Phys.: Condens. Matter}\
  }\textbf {\bibinfo {volume} {16}},\ \bibinfo {pages} {S497} (\bibinfo {year}
  {2004})}\BibitemShut {NoStop}%
\bibitem [{\citenamefont {Klauss}\ \emph {et~al.}(2008)\citenamefont {Klauss},
  \citenamefont {Luetkens}, \citenamefont {Klingeler}, \citenamefont {Hess},
  \citenamefont {Litterst}, \citenamefont {Kraken}, \citenamefont {Korshunov},
  \citenamefont {Eremin}, \citenamefont {Drechsler}, \citenamefont {Khasanov},
  \citenamefont {Amato}, \citenamefont {Hamann-Borrero}, \citenamefont {Leps},
  \citenamefont {Kondrat}, \citenamefont {Behr}, \citenamefont {Werner},\ and\
  \citenamefont {B\"uchner}}]{PhysRevLett.101.077005}%
  \BibitemOpen
  \bibfield  {author} {\bibinfo {author} {\bibfnamefont {H.-H.}\ \bibnamefont
  {Klauss}}, \bibinfo {author} {\bibfnamefont {H.}~\bibnamefont {Luetkens}},
  \bibinfo {author} {\bibfnamefont {R.}~\bibnamefont {Klingeler}}, \bibinfo
  {author} {\bibfnamefont {C.}~\bibnamefont {Hess}}, \bibinfo {author}
  {\bibfnamefont {F.~J.}\ \bibnamefont {Litterst}}, \bibinfo {author}
  {\bibfnamefont {M.}~\bibnamefont {Kraken}}, \bibinfo {author} {\bibfnamefont
  {M.~M.}\ \bibnamefont {Korshunov}}, \bibinfo {author} {\bibfnamefont
  {I.}~\bibnamefont {Eremin}}, \bibinfo {author} {\bibfnamefont {S.-L.}\
  \bibnamefont {Drechsler}}, \bibinfo {author} {\bibfnamefont {R.}~\bibnamefont
  {Khasanov}}, \bibinfo {author} {\bibfnamefont {A.}~\bibnamefont {Amato}},
  \bibinfo {author} {\bibfnamefont {J.}~\bibnamefont {Hamann-Borrero}},
  \bibinfo {author} {\bibfnamefont {N.}~\bibnamefont {Leps}}, \bibinfo {author}
  {\bibfnamefont {A.}~\bibnamefont {Kondrat}}, \bibinfo {author} {\bibfnamefont
  {G.}~\bibnamefont {Behr}}, \bibinfo {author} {\bibfnamefont {J.}~\bibnamefont
  {Werner}}, \ and\ \bibinfo {author} {\bibfnamefont {B.}~\bibnamefont
  {B\"uchner}},\ }\href {\doibase 10.1103/PhysRevLett.101.077005} {\bibfield
  {journal} {\bibinfo  {journal} {Phys. Rev. Lett.}\ }\textbf {\bibinfo
  {volume} {101}},\ \bibinfo {pages} {077005} (\bibinfo {year}
  {2008})}\BibitemShut {NoStop}%
\bibitem [{\citenamefont {McGuire}\ \emph {et~al.}(2008)\citenamefont
  {McGuire}, \citenamefont {Christianson}, \citenamefont {Sefat}, \citenamefont
  {Sales}, \citenamefont {Lumsden}, \citenamefont {Jin}, \citenamefont
  {Payzant}, \citenamefont {Mandrus}, \citenamefont {Luan}, \citenamefont
  {Keppens}, \citenamefont {Varadarajan}, \citenamefont {Brill}, \citenamefont
  {Hermann}, \citenamefont {Sougrati}, \citenamefont {Grandjean},\ and\
  \citenamefont {Long}}]{PhysRevB.78.094517}%
  \BibitemOpen
  \bibfield  {author} {\bibinfo {author} {\bibfnamefont {M.~A.}\ \bibnamefont
  {McGuire}}, \bibinfo {author} {\bibfnamefont {A.~D.}\ \bibnamefont
  {Christianson}}, \bibinfo {author} {\bibfnamefont {A.~S.}\ \bibnamefont
  {Sefat}}, \bibinfo {author} {\bibfnamefont {B.~C.}\ \bibnamefont {Sales}},
  \bibinfo {author} {\bibfnamefont {M.~D.}\ \bibnamefont {Lumsden}}, \bibinfo
  {author} {\bibfnamefont {R.}~\bibnamefont {Jin}}, \bibinfo {author}
  {\bibfnamefont {E.~A.}\ \bibnamefont {Payzant}}, \bibinfo {author}
  {\bibfnamefont {D.}~\bibnamefont {Mandrus}}, \bibinfo {author} {\bibfnamefont
  {Y.}~\bibnamefont {Luan}}, \bibinfo {author} {\bibfnamefont {V.}~\bibnamefont
  {Keppens}}, \bibinfo {author} {\bibfnamefont {V.}~\bibnamefont
  {Varadarajan}}, \bibinfo {author} {\bibfnamefont {J.~W.}\ \bibnamefont
  {Brill}}, \bibinfo {author} {\bibfnamefont {R.~P.}\ \bibnamefont {Hermann}},
  \bibinfo {author} {\bibfnamefont {M.~T.}\ \bibnamefont {Sougrati}}, \bibinfo
  {author} {\bibfnamefont {F.}~\bibnamefont {Grandjean}}, \ and\ \bibinfo
  {author} {\bibfnamefont {G.~J.}\ \bibnamefont {Long}},\ }\href {\doibase
  10.1103/PhysRevB.78.094517} {\bibfield  {journal} {\bibinfo  {journal} {Phys.
  Rev. B}\ }\textbf {\bibinfo {volume} {78}},\ \bibinfo {pages} {094517}
  (\bibinfo {year} {2008})}\BibitemShut {NoStop}%
\bibitem [{\citenamefont {Materne}\ \emph {et~al.}(2018)\citenamefont
  {Materne}, \citenamefont {Bi}, \citenamefont {Alp}, \citenamefont {Zhao},
  \citenamefont {Hu}, \citenamefont {Jesche}, \citenamefont {Geibel},
  \citenamefont {Kappenberger}, \citenamefont {Aswartham}, \citenamefont
  {Wurmehl}, \citenamefont {B\"uchner}, \citenamefont {Zhang}, \citenamefont
  {Goltz}, \citenamefont {Spehling},\ and\ \citenamefont
  {Klauss}}]{PhysRevB.98.014517}%
  \BibitemOpen
  \bibfield  {author} {\bibinfo {author} {\bibfnamefont {P.}~\bibnamefont
  {Materne}}, \bibinfo {author} {\bibfnamefont {W.}~\bibnamefont {Bi}},
  \bibinfo {author} {\bibfnamefont {E.~E.}\ \bibnamefont {Alp}}, \bibinfo
  {author} {\bibfnamefont {J.}~\bibnamefont {Zhao}}, \bibinfo {author}
  {\bibfnamefont {M.~Y.}\ \bibnamefont {Hu}}, \bibinfo {author} {\bibfnamefont
  {A.}~\bibnamefont {Jesche}}, \bibinfo {author} {\bibfnamefont
  {C.}~\bibnamefont {Geibel}}, \bibinfo {author} {\bibfnamefont
  {R.}~\bibnamefont {Kappenberger}}, \bibinfo {author} {\bibfnamefont
  {S.}~\bibnamefont {Aswartham}}, \bibinfo {author} {\bibfnamefont
  {S.}~\bibnamefont {Wurmehl}}, \bibinfo {author} {\bibfnamefont
  {B.}~\bibnamefont {B\"uchner}}, \bibinfo {author} {\bibfnamefont
  {D.}~\bibnamefont {Zhang}}, \bibinfo {author} {\bibfnamefont
  {T.}~\bibnamefont {Goltz}}, \bibinfo {author} {\bibfnamefont
  {J.}~\bibnamefont {Spehling}}, \ and\ \bibinfo {author} {\bibfnamefont
  {H.-H.}\ \bibnamefont {Klauss}},\ }\href {\doibase
  10.1103/PhysRevB.98.014517} {\bibfield  {journal} {\bibinfo  {journal} {Phys.
  Rev. B}\ }\textbf {\bibinfo {volume} {98}},\ \bibinfo {pages} {014517}
  (\bibinfo {year} {2018})}\BibitemShut {NoStop}%
\bibitem [{\citenamefont {B\l{}achowski}\ \emph {et~al.}(2010)\citenamefont
  {B\l{}achowski}, \citenamefont {Ruebenbauer}, \citenamefont {Zukrowski},
  \citenamefont {Przewoznik}, \citenamefont {Wojciechowski},\ and\
  \citenamefont {Stadnik}}]{Blachowski20101}%
  \BibitemOpen
  \bibfield  {author} {\bibinfo {author} {\bibfnamefont {A.}~\bibnamefont
  {B\l{}achowski}}, \bibinfo {author} {\bibfnamefont {K.}~\bibnamefont
  {Ruebenbauer}}, \bibinfo {author} {\bibfnamefont {J.}~\bibnamefont
  {Zukrowski}}, \bibinfo {author} {\bibfnamefont {J.}~\bibnamefont
  {Przewoznik}}, \bibinfo {author} {\bibfnamefont {K.}~\bibnamefont
  {Wojciechowski}}, \ and\ \bibinfo {author} {\bibfnamefont {Z.}~\bibnamefont
  {Stadnik}},\ }\href {\doibase 10.1016/j.jallcom.2009.12.095} {\bibfield
  {journal} {\bibinfo  {journal} {J. Alloys Compd.}\ }\textbf {\bibinfo
  {volume} {494}},\ \bibinfo {pages} {1 } (\bibinfo {year} {2010})}\BibitemShut
  {NoStop}%
\bibitem [{\citenamefont {Renzi}\ \emph {et~al.}(2012)\citenamefont {Renzi},
  \citenamefont {Bonfà}, \citenamefont {Mazzani}, \citenamefont {Sanna},
  \citenamefont {Prando}, \citenamefont {Carretta}, \citenamefont {Khasanov},
  \citenamefont {Amato}, \citenamefont {Luetkens}, \citenamefont {Bendele},
  \citenamefont {Bernardini}, \citenamefont {Massidda}, \citenamefont
  {Palenzona}, \citenamefont {Tropeano},\ and\ \citenamefont
  {Vignolo}}]{0953-2048-25-8-084009}%
  \BibitemOpen
  \bibfield  {author} {\bibinfo {author} {\bibfnamefont {R.~D.}\ \bibnamefont
  {Renzi}}, \bibinfo {author} {\bibfnamefont {P.}~\bibnamefont {Bonfà}},
  \bibinfo {author} {\bibfnamefont {M.}~\bibnamefont {Mazzani}}, \bibinfo
  {author} {\bibfnamefont {S.}~\bibnamefont {Sanna}}, \bibinfo {author}
  {\bibfnamefont {G.}~\bibnamefont {Prando}}, \bibinfo {author} {\bibfnamefont
  {P.}~\bibnamefont {Carretta}}, \bibinfo {author} {\bibfnamefont
  {R.}~\bibnamefont {Khasanov}}, \bibinfo {author} {\bibfnamefont
  {A.}~\bibnamefont {Amato}}, \bibinfo {author} {\bibfnamefont
  {H.}~\bibnamefont {Luetkens}}, \bibinfo {author} {\bibfnamefont
  {M.}~\bibnamefont {Bendele}}, \bibinfo {author} {\bibfnamefont
  {F.}~\bibnamefont {Bernardini}}, \bibinfo {author} {\bibfnamefont
  {S.}~\bibnamefont {Massidda}}, \bibinfo {author} {\bibfnamefont
  {A.}~\bibnamefont {Palenzona}}, \bibinfo {author} {\bibfnamefont
  {M.}~\bibnamefont {Tropeano}}, \ and\ \bibinfo {author} {\bibfnamefont
  {M.}~\bibnamefont {Vignolo}},\ }\href
  {http://stacks.iop.org/0953-2048/25/i=8/a=084009} {\bibfield  {journal}
  {\bibinfo  {journal} {Supercond. Sci. Technol.}\ }\textbf {\bibinfo {volume}
  {25}},\ \bibinfo {pages} {084009} (\bibinfo {year} {2012})}\BibitemShut
  {NoStop}%
\bibitem [{\citenamefont {Materne}\ \emph {et~al.}(2015)\citenamefont
  {Materne}, \citenamefont {Kamusella}, \citenamefont {Sarkar}, \citenamefont
  {Goltz}, \citenamefont {Spehling}, \citenamefont {Maeter}, \citenamefont
  {Harnagea}, \citenamefont {Wurmehl}, \citenamefont {B\"uchner}, \citenamefont
  {Luetkens}, \citenamefont {Timm},\ and\ \citenamefont
  {Klauss}}]{PhysRevB.92.134511}%
  \BibitemOpen
  \bibfield  {author} {\bibinfo {author} {\bibfnamefont {P.}~\bibnamefont
  {Materne}}, \bibinfo {author} {\bibfnamefont {S.}~\bibnamefont {Kamusella}},
  \bibinfo {author} {\bibfnamefont {R.}~\bibnamefont {Sarkar}}, \bibinfo
  {author} {\bibfnamefont {T.}~\bibnamefont {Goltz}}, \bibinfo {author}
  {\bibfnamefont {J.}~\bibnamefont {Spehling}}, \bibinfo {author}
  {\bibfnamefont {H.}~\bibnamefont {Maeter}}, \bibinfo {author} {\bibfnamefont
  {L.}~\bibnamefont {Harnagea}}, \bibinfo {author} {\bibfnamefont
  {S.}~\bibnamefont {Wurmehl}}, \bibinfo {author} {\bibfnamefont
  {B.}~\bibnamefont {B\"uchner}}, \bibinfo {author} {\bibfnamefont
  {H.}~\bibnamefont {Luetkens}}, \bibinfo {author} {\bibfnamefont
  {C.}~\bibnamefont {Timm}}, \ and\ \bibinfo {author} {\bibfnamefont {H.-H.}\
  \bibnamefont {Klauss}},\ }\href {\doibase 10.1103/PhysRevB.92.134511}
  {\bibfield  {journal} {\bibinfo  {journal} {Phys. Rev. B}\ }\textbf {\bibinfo
  {volume} {92}},\ \bibinfo {pages} {134511} (\bibinfo {year}
  {2015})}\BibitemShut {NoStop}%
\bibitem [{\citenamefont {Goltz}\ \emph {et~al.}(2014)\citenamefont {Goltz},
  \citenamefont {Zinth}, \citenamefont {Johrendt}, \citenamefont {Rosner},
  \citenamefont {Pascua}, \citenamefont {Luetkens}, \citenamefont {Materne},\
  and\ \citenamefont {Klauss}}]{PhysRevB.89.144511}%
  \BibitemOpen
  \bibfield  {author} {\bibinfo {author} {\bibfnamefont {T.}~\bibnamefont
  {Goltz}}, \bibinfo {author} {\bibfnamefont {V.}~\bibnamefont {Zinth}},
  \bibinfo {author} {\bibfnamefont {D.}~\bibnamefont {Johrendt}}, \bibinfo
  {author} {\bibfnamefont {H.}~\bibnamefont {Rosner}}, \bibinfo {author}
  {\bibfnamefont {G.}~\bibnamefont {Pascua}}, \bibinfo {author} {\bibfnamefont
  {H.}~\bibnamefont {Luetkens}}, \bibinfo {author} {\bibfnamefont
  {P.}~\bibnamefont {Materne}}, \ and\ \bibinfo {author} {\bibfnamefont
  {H.-H.}\ \bibnamefont {Klauss}},\ }\href {\doibase
  10.1103/PhysRevB.89.144511} {\bibfield  {journal} {\bibinfo  {journal} {Phys.
  Rev. B}\ }\textbf {\bibinfo {volume} {89}},\ \bibinfo {pages} {144511}
  (\bibinfo {year} {2014})}\BibitemShut {NoStop}%
\bibitem [{\citenamefont {Prando}\ \emph {et~al.}(2013)\citenamefont {Prando},
  \citenamefont {Vakaliuk}, \citenamefont {Sanna}, \citenamefont {Lamura},
  \citenamefont {Shiroka}, \citenamefont {Bonf\`a}, \citenamefont {Carretta},
  \citenamefont {De~Renzi}, \citenamefont {Klauss}, \citenamefont {Blum},
  \citenamefont {Wurmehl}, \citenamefont {Hess},\ and\ \citenamefont
  {B\"uchner}}]{PhysRevB.87.174519}%
  \BibitemOpen
  \bibfield  {author} {\bibinfo {author} {\bibfnamefont {G.}~\bibnamefont
  {Prando}}, \bibinfo {author} {\bibfnamefont {O.}~\bibnamefont {Vakaliuk}},
  \bibinfo {author} {\bibfnamefont {S.}~\bibnamefont {Sanna}}, \bibinfo
  {author} {\bibfnamefont {G.}~\bibnamefont {Lamura}}, \bibinfo {author}
  {\bibfnamefont {T.}~\bibnamefont {Shiroka}}, \bibinfo {author} {\bibfnamefont
  {P.}~\bibnamefont {Bonf\`a}}, \bibinfo {author} {\bibfnamefont
  {P.}~\bibnamefont {Carretta}}, \bibinfo {author} {\bibfnamefont
  {R.}~\bibnamefont {De~Renzi}}, \bibinfo {author} {\bibfnamefont {H.-H.}\
  \bibnamefont {Klauss}}, \bibinfo {author} {\bibfnamefont {C.~G.~F.}\
  \bibnamefont {Blum}}, \bibinfo {author} {\bibfnamefont {S.}~\bibnamefont
  {Wurmehl}}, \bibinfo {author} {\bibfnamefont {C.}~\bibnamefont {Hess}}, \
  and\ \bibinfo {author} {\bibfnamefont {B.}~\bibnamefont {B\"uchner}},\ }\href
  {\doibase 10.1103/PhysRevB.87.174519} {\bibfield  {journal} {\bibinfo
  {journal} {Phys. Rev. B}\ }\textbf {\bibinfo {volume} {87}},\ \bibinfo
  {pages} {174519} (\bibinfo {year} {2013})}\BibitemShut {NoStop}%
\bibitem [{\citenamefont {Ricci}\ \emph {et~al.}(2010)\citenamefont {Ricci},
  \citenamefont {Poccia}, \citenamefont {Joseph}, \citenamefont {Barba},
  \citenamefont {Arrighetti}, \citenamefont {Ciasca}, \citenamefont {Yan},
  \citenamefont {McCallum}, \citenamefont {Lograsso}, \citenamefont {Zhigadlo},
  \citenamefont {Karpinski},\ and\ \citenamefont
  {Bianconi}}]{PhysRevB.82.144507}%
  \BibitemOpen
  \bibfield  {author} {\bibinfo {author} {\bibfnamefont {A.}~\bibnamefont
  {Ricci}}, \bibinfo {author} {\bibfnamefont {N.}~\bibnamefont {Poccia}},
  \bibinfo {author} {\bibfnamefont {B.}~\bibnamefont {Joseph}}, \bibinfo
  {author} {\bibfnamefont {L.}~\bibnamefont {Barba}}, \bibinfo {author}
  {\bibfnamefont {G.}~\bibnamefont {Arrighetti}}, \bibinfo {author}
  {\bibfnamefont {G.}~\bibnamefont {Ciasca}}, \bibinfo {author} {\bibfnamefont
  {J.-Q.}\ \bibnamefont {Yan}}, \bibinfo {author} {\bibfnamefont {R.~W.}\
  \bibnamefont {McCallum}}, \bibinfo {author} {\bibfnamefont {T.~A.}\
  \bibnamefont {Lograsso}}, \bibinfo {author} {\bibfnamefont {N.~D.}\
  \bibnamefont {Zhigadlo}}, \bibinfo {author} {\bibfnamefont {J.}~\bibnamefont
  {Karpinski}}, \ and\ \bibinfo {author} {\bibfnamefont {A.}~\bibnamefont
  {Bianconi}},\ }\href {\doibase 10.1103/PhysRevB.82.144507} {\bibfield
  {journal} {\bibinfo  {journal} {Phys. Rev. B}\ }\textbf {\bibinfo {volume}
  {82}},\ \bibinfo {pages} {144507} (\bibinfo {year} {2010})}\BibitemShut
  {NoStop}%
\bibitem [{\citenamefont {Kobayashi}\ \emph {et~al.}(2016)\citenamefont
  {Kobayashi}, \citenamefont {Yamaura}, \citenamefont {Iimura}, \citenamefont
  {Maki}, \citenamefont {Sagayama}, \citenamefont {Kumai}, \citenamefont
  {Murakami}, \citenamefont {Takahashi}, \citenamefont {Matsuishi},\ and\
  \citenamefont {Hosono}}]{Kobayashi2016}%
  \BibitemOpen
  \bibfield  {author} {\bibinfo {author} {\bibfnamefont {K.}~\bibnamefont
  {Kobayashi}}, \bibinfo {author} {\bibfnamefont {J.-i.}\ \bibnamefont
  {Yamaura}}, \bibinfo {author} {\bibfnamefont {S.}~\bibnamefont {Iimura}},
  \bibinfo {author} {\bibfnamefont {S.}~\bibnamefont {Maki}}, \bibinfo {author}
  {\bibfnamefont {H.}~\bibnamefont {Sagayama}}, \bibinfo {author}
  {\bibfnamefont {R.}~\bibnamefont {Kumai}}, \bibinfo {author} {\bibfnamefont
  {Y.}~\bibnamefont {Murakami}}, \bibinfo {author} {\bibfnamefont
  {H.}~\bibnamefont {Takahashi}}, \bibinfo {author} {\bibfnamefont
  {S.}~\bibnamefont {Matsuishi}}, \ and\ \bibinfo {author} {\bibfnamefont
  {H.}~\bibnamefont {Hosono}},\ }\href {\doibase 10.1038/srep39646} {\bibfield
  {journal} {\bibinfo  {journal} {Sci. Rep.}\ }\textbf {\bibinfo {volume}
  {6}},\ \bibinfo {pages} {39646} (\bibinfo {year} {2016})}\BibitemShut
  {NoStop}%
\bibitem [{\citenamefont {Mirbt}\ \emph {et~al.}(2003)\citenamefont {Mirbt},
  \citenamefont {Sanyal}, \citenamefont {Isheden},\ and\ \citenamefont
  {Johansson}}]{PhysRevB.67.155421}%
  \BibitemOpen
  \bibfield  {author} {\bibinfo {author} {\bibfnamefont {S.}~\bibnamefont
  {Mirbt}}, \bibinfo {author} {\bibfnamefont {B.}~\bibnamefont {Sanyal}},
  \bibinfo {author} {\bibfnamefont {C.}~\bibnamefont {Isheden}}, \ and\
  \bibinfo {author} {\bibfnamefont {B.}~\bibnamefont {Johansson}},\ }\href
  {\doibase 10.1103/PhysRevB.67.155421} {\bibfield  {journal} {\bibinfo
  {journal} {Phys. Rev. B}\ }\textbf {\bibinfo {volume} {67}},\ \bibinfo
  {pages} {155421} (\bibinfo {year} {2003})}\BibitemShut {NoStop}%
\bibitem [{\citenamefont {McQueen}\ \emph {et~al.}(2008)\citenamefont
  {McQueen}, \citenamefont {Regulacio}, \citenamefont {Williams}, \citenamefont
  {Huang}, \citenamefont {Lynn}, \citenamefont {Hor}, \citenamefont {West},
  \citenamefont {Green},\ and\ \citenamefont {Cava}}]{PhysRevB.78.024521}%
  \BibitemOpen
  \bibfield  {author} {\bibinfo {author} {\bibfnamefont {T.~M.}\ \bibnamefont
  {McQueen}}, \bibinfo {author} {\bibfnamefont {M.}~\bibnamefont {Regulacio}},
  \bibinfo {author} {\bibfnamefont {A.~J.}\ \bibnamefont {Williams}}, \bibinfo
  {author} {\bibfnamefont {Q.}~\bibnamefont {Huang}}, \bibinfo {author}
  {\bibfnamefont {J.~W.}\ \bibnamefont {Lynn}}, \bibinfo {author}
  {\bibfnamefont {Y.~S.}\ \bibnamefont {Hor}}, \bibinfo {author} {\bibfnamefont
  {D.~V.}\ \bibnamefont {West}}, \bibinfo {author} {\bibfnamefont {M.~A.}\
  \bibnamefont {Green}}, \ and\ \bibinfo {author} {\bibfnamefont {R.~J.}\
  \bibnamefont {Cava}},\ }\href {\doibase 10.1103/PhysRevB.78.024521}
  {\bibfield  {journal} {\bibinfo  {journal} {Phys. Rev. B}\ }\textbf {\bibinfo
  {volume} {78}},\ \bibinfo {pages} {024521} (\bibinfo {year}
  {2008})}\BibitemShut {NoStop}%
\bibitem [{\citenamefont {Wu}\ \emph {et~al.}(2008)\citenamefont {Wu},
  \citenamefont {Phillips},\ and\ \citenamefont
  {Castro~Neto}}]{PhysRevLett.101.126401}%
  \BibitemOpen
  \bibfield  {author} {\bibinfo {author} {\bibfnamefont {J.}~\bibnamefont
  {Wu}}, \bibinfo {author} {\bibfnamefont {P.}~\bibnamefont {Phillips}}, \ and\
  \bibinfo {author} {\bibfnamefont {A.~H.}\ \bibnamefont {Castro~Neto}},\
  }\href {\doibase 10.1103/PhysRevLett.101.126401} {\bibfield  {journal}
  {\bibinfo  {journal} {Phys. Rev. Lett.}\ }\textbf {\bibinfo {volume} {101}},\
  \bibinfo {pages} {126401} (\bibinfo {year} {2008})}\BibitemShut {NoStop}%
\bibitem [{\citenamefont {Opahle}\ \emph {et~al.}(2009)\citenamefont {Opahle},
  \citenamefont {Kandpal}, \citenamefont {Zhang}, \citenamefont {Gros},\ and\
  \citenamefont {Valent\'{\i}}}]{PhysRevB.79.024509}%
  \BibitemOpen
  \bibfield  {author} {\bibinfo {author} {\bibfnamefont {I.}~\bibnamefont
  {Opahle}}, \bibinfo {author} {\bibfnamefont {H.~C.}\ \bibnamefont {Kandpal}},
  \bibinfo {author} {\bibfnamefont {Y.}~\bibnamefont {Zhang}}, \bibinfo
  {author} {\bibfnamefont {C.}~\bibnamefont {Gros}}, \ and\ \bibinfo {author}
  {\bibfnamefont {R.}~\bibnamefont {Valent\'{\i}}},\ }\href {\doibase
  10.1103/PhysRevB.79.024509} {\bibfield  {journal} {\bibinfo  {journal} {Phys.
  Rev. B}\ }\textbf {\bibinfo {volume} {79}},\ \bibinfo {pages} {024509}
  (\bibinfo {year} {2009})}\BibitemShut {NoStop}%
\bibitem [{\citenamefont {McElroy}\ \emph {et~al.}(2013)\citenamefont
  {McElroy}, \citenamefont {Hamlin}, \citenamefont {White}, \citenamefont
  {McGuire}, \citenamefont {Sales},\ and\ \citenamefont
  {Maple}}]{PhysRevB.88.134513}%
  \BibitemOpen
  \bibfield  {author} {\bibinfo {author} {\bibfnamefont {C.~A.}\ \bibnamefont
  {McElroy}}, \bibinfo {author} {\bibfnamefont {J.~J.}\ \bibnamefont {Hamlin}},
  \bibinfo {author} {\bibfnamefont {B.~D.}\ \bibnamefont {White}}, \bibinfo
  {author} {\bibfnamefont {M.~A.}\ \bibnamefont {McGuire}}, \bibinfo {author}
  {\bibfnamefont {B.~C.}\ \bibnamefont {Sales}}, \ and\ \bibinfo {author}
  {\bibfnamefont {M.~B.}\ \bibnamefont {Maple}},\ }\href {\doibase
  10.1103/PhysRevB.88.134513} {\bibfield  {journal} {\bibinfo  {journal} {Phys.
  Rev. B}\ }\textbf {\bibinfo {volume} {88}},\ \bibinfo {pages} {134513}
  (\bibinfo {year} {2013})}\BibitemShut {NoStop}%
\bibitem [{\citenamefont {Jesche}\ \emph {et~al.}(2012)\citenamefont {Jesche},
  \citenamefont {Nitsche}, \citenamefont {Probst}, \citenamefont {Doert},
  \citenamefont {M\"uller},\ and\ \citenamefont {Ruck}}]{PhysRevB.86.134511}%
  \BibitemOpen
  \bibfield  {author} {\bibinfo {author} {\bibfnamefont {A.}~\bibnamefont
  {Jesche}}, \bibinfo {author} {\bibfnamefont {F.}~\bibnamefont {Nitsche}},
  \bibinfo {author} {\bibfnamefont {S.}~\bibnamefont {Probst}}, \bibinfo
  {author} {\bibfnamefont {T.}~\bibnamefont {Doert}}, \bibinfo {author}
  {\bibfnamefont {P.}~\bibnamefont {M\"uller}}, \ and\ \bibinfo {author}
  {\bibfnamefont {M.}~\bibnamefont {Ruck}},\ }\href {\doibase
  10.1103/PhysRevB.86.134511} {\bibfield  {journal} {\bibinfo  {journal} {Phys.
  Rev. B}\ }\textbf {\bibinfo {volume} {86}},\ \bibinfo {pages} {134511}
  (\bibinfo {year} {2012})}\BibitemShut {NoStop}%
\bibitem [{\citenamefont {Raffius}\ \emph {et~al.}(1993)\citenamefont
  {Raffius}, \citenamefont {Mörsen}, \citenamefont {Mosel}, \citenamefont
  {Müller-Warmuth}, \citenamefont {Jeitschko}, \citenamefont {Terbüchte},\ and\
  \citenamefont {Vomhof}}]{RAFFIUS1993135}%
  \BibitemOpen
  \bibfield  {author} {\bibinfo {author} {\bibfnamefont {H.}~\bibnamefont
  {Raffius}}, \bibinfo {author} {\bibfnamefont {E.}~\bibnamefont {Mörsen}},
  \bibinfo {author} {\bibfnamefont {B.}~\bibnamefont {Mosel}}, \bibinfo
  {author} {\bibfnamefont {W.}~\bibnamefont {Müller-Warmuth}}, \bibinfo
  {author} {\bibfnamefont {W.}~\bibnamefont {Jeitschko}}, \bibinfo {author}
  {\bibfnamefont {L.}~\bibnamefont {Terbüchte}}, \ and\ \bibinfo {author}
  {\bibfnamefont {T.}~\bibnamefont {Vomhof}},\ }\href {\doibase
  https://doi.org/10.1016/0022-3697(93)90301-7} {\bibfield  {journal} {\bibinfo
   {journal} {J. Phys. Chem. Solids}\ }\textbf {\bibinfo {volume} {54}},\
  \bibinfo {pages} {135 } (\bibinfo {year} {1993})}\BibitemShut {NoStop}%
\bibitem [{\citenamefont {Okada}\ \emph {et~al.}(2008)\citenamefont {Okada},
  \citenamefont {Igawa}, \citenamefont {Takahashi}, \citenamefont {Kamihara},
  \citenamefont {Hirano}, \citenamefont {Hosono}, \citenamefont
  {Matsubayashi},\ and\ \citenamefont {Uwatoko}}]{JPSJ.77.113712}%
  \BibitemOpen
  \bibfield  {author} {\bibinfo {author} {\bibfnamefont {H.}~\bibnamefont
  {Okada}}, \bibinfo {author} {\bibfnamefont {K.}~\bibnamefont {Igawa}},
  \bibinfo {author} {\bibfnamefont {H.}~\bibnamefont {Takahashi}}, \bibinfo
  {author} {\bibfnamefont {Y.}~\bibnamefont {Kamihara}}, \bibinfo {author}
  {\bibfnamefont {M.}~\bibnamefont {Hirano}}, \bibinfo {author} {\bibfnamefont
  {H.}~\bibnamefont {Hosono}}, \bibinfo {author} {\bibfnamefont
  {K.}~\bibnamefont {Matsubayashi}}, \ and\ \bibinfo {author} {\bibfnamefont
  {Y.}~\bibnamefont {Uwatoko}},\ }\href {\doibase 10.1143/JPSJ.77.113712}
  {\bibfield  {journal} {\bibinfo  {journal} {J. Phys. Soc. Jpn.}\ }\textbf
  {\bibinfo {volume} {77}},\ \bibinfo {pages} {113712} (\bibinfo {year}
  {2008})}\BibitemShut {NoStop}%
\bibitem [{\citenamefont {Qureshi}\ \emph {et~al.}(2010)\citenamefont
  {Qureshi}, \citenamefont {Drees}, \citenamefont {Werner}, \citenamefont
  {Wurmehl}, \citenamefont {Hess}, \citenamefont {Klingeler}, \citenamefont
  {B\"uchner}, \citenamefont {Fern\'andez-D\'iaz},\ and\ \citenamefont
  {Braden}}]{PhysRevB.82.184521}%
  \BibitemOpen
  \bibfield  {author} {\bibinfo {author} {\bibfnamefont {N.}~\bibnamefont
  {Qureshi}}, \bibinfo {author} {\bibfnamefont {Y.}~\bibnamefont {Drees}},
  \bibinfo {author} {\bibfnamefont {J.}~\bibnamefont {Werner}}, \bibinfo
  {author} {\bibfnamefont {S.}~\bibnamefont {Wurmehl}}, \bibinfo {author}
  {\bibfnamefont {C.}~\bibnamefont {Hess}}, \bibinfo {author} {\bibfnamefont
  {R.}~\bibnamefont {Klingeler}}, \bibinfo {author} {\bibfnamefont
  {B.}~\bibnamefont {B\"uchner}}, \bibinfo {author} {\bibfnamefont {M.~T.}\
  \bibnamefont {Fern\'andez-D\'iaz}}, \ and\ \bibinfo {author} {\bibfnamefont
  {M.}~\bibnamefont {Braden}},\ }\href {\doibase 10.1103/PhysRevB.82.184521}
  {\bibfield  {journal} {\bibinfo  {journal} {Phys. Rev. B}\ }\textbf {\bibinfo
  {volume} {82}},\ \bibinfo {pages} {184521} (\bibinfo {year}
  {2010})}\BibitemShut {NoStop}%
\bibitem [{\citenamefont {Nomura}\ \emph {et~al.}(2008)\citenamefont {Nomura},
  \citenamefont {Kim}, \citenamefont {Kamihara}, \citenamefont {Hirano},
  \citenamefont {Sushko}, \citenamefont {Kato}, \citenamefont {Takata},
  \citenamefont {Shluger},\ and\ \citenamefont
  {Hosono}}]{0953-2048-21-12-125028}%
  \BibitemOpen
  \bibfield  {author} {\bibinfo {author} {\bibfnamefont {T.}~\bibnamefont
  {Nomura}}, \bibinfo {author} {\bibfnamefont {S.~W.}\ \bibnamefont {Kim}},
  \bibinfo {author} {\bibfnamefont {Y.}~\bibnamefont {Kamihara}}, \bibinfo
  {author} {\bibfnamefont {M.}~\bibnamefont {Hirano}}, \bibinfo {author}
  {\bibfnamefont {P.~V.}\ \bibnamefont {Sushko}}, \bibinfo {author}
  {\bibfnamefont {K.}~\bibnamefont {Kato}}, \bibinfo {author} {\bibfnamefont
  {M.}~\bibnamefont {Takata}}, \bibinfo {author} {\bibfnamefont {A.~L.}\
  \bibnamefont {Shluger}}, \ and\ \bibinfo {author} {\bibfnamefont
  {H.}~\bibnamefont {Hosono}},\ }\href
  {http://stacks.iop.org/0953-2048/21/i=12/a=125028} {\bibfield  {journal}
  {\bibinfo  {journal} {Supercond. Sci. Technol.}\ }\textbf {\bibinfo {volume}
  {21}},\ \bibinfo {pages} {125028} (\bibinfo {year} {2008})}\BibitemShut
  {NoStop}%
\bibitem [{\citenamefont {Andersen}\ and\ \citenamefont
  {Boeri}(2011)}]{ANDP201000149}%
  \BibitemOpen
  \bibfield  {author} {\bibinfo {author} {\bibfnamefont {O.}~\bibnamefont
  {Andersen}}\ and\ \bibinfo {author} {\bibfnamefont {L.}~\bibnamefont
  {Boeri}},\ }\href {\doibase 10.1002/andp.201000149} {\bibfield  {journal}
  {\bibinfo  {journal} {Ann. Phys.}\ }\textbf {\bibinfo {volume} {523}},\
  \bibinfo {pages} {8} (\bibinfo {year} {2011})}\BibitemShut {NoStop}%
\bibitem [{\citenamefont {Xu}\ \emph {et~al.}(2008)\citenamefont {Xu},
  \citenamefont {M\"uller},\ and\ \citenamefont
  {Sachdev}}]{PhysRevB.78.020501}%
  \BibitemOpen
  \bibfield  {author} {\bibinfo {author} {\bibfnamefont {C.}~\bibnamefont
  {Xu}}, \bibinfo {author} {\bibfnamefont {M.}~\bibnamefont {M\"uller}}, \ and\
  \bibinfo {author} {\bibfnamefont {S.}~\bibnamefont {Sachdev}},\ }\href
  {\doibase 10.1103/PhysRevB.78.020501} {\bibfield  {journal} {\bibinfo
  {journal} {Phys. Rev. B}\ }\textbf {\bibinfo {volume} {78}},\ \bibinfo
  {pages} {020501} (\bibinfo {year} {2008})}\BibitemShut {NoStop}%
\bibitem [{\citenamefont {Kou}\ \emph {et~al.}(2009)\citenamefont {Kou},
  \citenamefont {Li},\ and\ \citenamefont {Weng}}]{0295-5075-88-1-17010}%
  \BibitemOpen
  \bibfield  {author} {\bibinfo {author} {\bibfnamefont {S.-P.}\ \bibnamefont
  {Kou}}, \bibinfo {author} {\bibfnamefont {T.}~\bibnamefont {Li}}, \ and\
  \bibinfo {author} {\bibfnamefont {Z.-Y.}\ \bibnamefont {Weng}},\ }\href
  {http://stacks.iop.org/0295-5075/88/i=1/a=17010} {\bibfield  {journal}
  {\bibinfo  {journal} {Europhys. Lett.}\ }\textbf {\bibinfo {volume} {88}},\
  \bibinfo {pages} {17010} (\bibinfo {year} {2009})}\BibitemShut {NoStop}%
\bibitem [{\citenamefont {Nowik}\ \emph {et~al.}(2008)\citenamefont {Nowik},
  \citenamefont {Felner}, \citenamefont {Awana}, \citenamefont {Vajpayee},\
  and\ \citenamefont {Kishan}}]{0953-8984-20-29-292201}%
  \BibitemOpen
  \bibfield  {author} {\bibinfo {author} {\bibfnamefont {I.}~\bibnamefont
  {Nowik}}, \bibinfo {author} {\bibfnamefont {I.}~\bibnamefont {Felner}},
  \bibinfo {author} {\bibfnamefont {V.~P.~S.}\ \bibnamefont {Awana}}, \bibinfo
  {author} {\bibfnamefont {A.}~\bibnamefont {Vajpayee}}, \ and\ \bibinfo
  {author} {\bibfnamefont {H.}~\bibnamefont {Kishan}},\ }\href
  {http://stacks.iop.org/0953-8984/20/i=29/a=292201} {\bibfield  {journal}
  {\bibinfo  {journal} {J. Phys.: Condens. Matter}\ }\textbf {\bibinfo {volume}
  {20}},\ \bibinfo {pages} {292201} (\bibinfo {year} {2008})}\BibitemShut
  {NoStop}%
\bibitem [{\citenamefont {Grinenko}\ \emph {et~al.}(2011)\citenamefont
  {Grinenko}, \citenamefont {Kikoin}, \citenamefont {Drechsler}, \citenamefont
  {Fuchs}, \citenamefont {Nenkov}, \citenamefont {Wurmehl}, \citenamefont
  {Hammerath}, \citenamefont {Lang}, \citenamefont {Grafe}, \citenamefont
  {Holzapfel}, \citenamefont {van~den Brink}, \citenamefont {B\"uchner},\ and\
  \citenamefont {Schultz}}]{PhysRevB.84.134516}%
  \BibitemOpen
  \bibfield  {author} {\bibinfo {author} {\bibfnamefont {V.}~\bibnamefont
  {Grinenko}}, \bibinfo {author} {\bibfnamefont {K.}~\bibnamefont {Kikoin}},
  \bibinfo {author} {\bibfnamefont {S.-L.}\ \bibnamefont {Drechsler}}, \bibinfo
  {author} {\bibfnamefont {G.}~\bibnamefont {Fuchs}}, \bibinfo {author}
  {\bibfnamefont {K.}~\bibnamefont {Nenkov}}, \bibinfo {author} {\bibfnamefont
  {S.}~\bibnamefont {Wurmehl}}, \bibinfo {author} {\bibfnamefont
  {F.}~\bibnamefont {Hammerath}}, \bibinfo {author} {\bibfnamefont
  {G.}~\bibnamefont {Lang}}, \bibinfo {author} {\bibfnamefont {H.-J.}\
  \bibnamefont {Grafe}}, \bibinfo {author} {\bibfnamefont {B.}~\bibnamefont
  {Holzapfel}}, \bibinfo {author} {\bibfnamefont {J.}~\bibnamefont {van~den
  Brink}}, \bibinfo {author} {\bibfnamefont {B.}~\bibnamefont {B\"uchner}}, \
  and\ \bibinfo {author} {\bibfnamefont {L.}~\bibnamefont {Schultz}},\ }\href
  {\doibase 10.1103/PhysRevB.84.134516} {\bibfield  {journal} {\bibinfo
  {journal} {Phys. Rev. B}\ }\textbf {\bibinfo {volume} {84}},\ \bibinfo
  {pages} {134516} (\bibinfo {year} {2011})}\BibitemShut {NoStop}%
\bibitem [{\citenamefont {Fu}\ \emph {et~al.}(2012)\citenamefont {Fu},
  \citenamefont {Torchetti}, \citenamefont {Imai}, \citenamefont {Ning},
  \citenamefont {Yan},\ and\ \citenamefont {Sefat}}]{PhysRevLett.109.247001}%
  \BibitemOpen
  \bibfield  {author} {\bibinfo {author} {\bibfnamefont {M.}~\bibnamefont
  {Fu}}, \bibinfo {author} {\bibfnamefont {D.~A.}\ \bibnamefont {Torchetti}},
  \bibinfo {author} {\bibfnamefont {T.}~\bibnamefont {Imai}}, \bibinfo {author}
  {\bibfnamefont {F.~L.}\ \bibnamefont {Ning}}, \bibinfo {author}
  {\bibfnamefont {J.-Q.}\ \bibnamefont {Yan}}, \ and\ \bibinfo {author}
  {\bibfnamefont {A.~S.}\ \bibnamefont {Sefat}},\ }\href {\doibase
  10.1103/PhysRevLett.109.247001} {\bibfield  {journal} {\bibinfo  {journal}
  {Phys. Rev. Lett.}\ }\textbf {\bibinfo {volume} {109}},\ \bibinfo {pages}
  {247001} (\bibinfo {year} {2012})}\BibitemShut {NoStop}%
\bibitem [{\citenamefont {Ok}\ \emph {et~al.}(2018)\citenamefont {Ok},
  \citenamefont {Baek}, \citenamefont {Efremov}, \citenamefont {Kappenberger},
  \citenamefont {Aswartham}, \citenamefont {Kim}, \citenamefont {van~den
  Brink},\ and\ \citenamefont {B\"uchner}}]{PhysRevB.97.180405}%
  \BibitemOpen
  \bibfield  {author} {\bibinfo {author} {\bibfnamefont {J.~M.}\ \bibnamefont
  {Ok}}, \bibinfo {author} {\bibfnamefont {S.-H.}\ \bibnamefont {Baek}},
  \bibinfo {author} {\bibfnamefont {D.~V.}\ \bibnamefont {Efremov}}, \bibinfo
  {author} {\bibfnamefont {R.}~\bibnamefont {Kappenberger}}, \bibinfo {author}
  {\bibfnamefont {S.}~\bibnamefont {Aswartham}}, \bibinfo {author}
  {\bibfnamefont {J.~S.}\ \bibnamefont {Kim}}, \bibinfo {author} {\bibfnamefont
  {J.}~\bibnamefont {van~den Brink}}, \ and\ \bibinfo {author} {\bibfnamefont
  {B.}~\bibnamefont {B\"uchner}},\ }\href {\doibase 10.1103/PhysRevB.97.180405}
  {\bibfield  {journal} {\bibinfo  {journal} {Phys. Rev. B}\ }\textbf {\bibinfo
  {volume} {97}},\ \bibinfo {pages} {180405} (\bibinfo {year}
  {2018})}\BibitemShut {NoStop}%
\end{thebibliography}%

\end{document}